\documentclass[fleqn,usenatbib]{new_mnras}

\usepackage{newtxtext,newtxmath}

\usepackage[T1]{fontenc}

\DeclareRobustCommand{\VAN}[3]{#2}
\let\VANthebibliography\thebibliography
\def\thebibliography{\DeclareRobustCommand{\VAN}[3]{##3}\VANthebibliography}

\usepackage{graphicx}

\newcommand{\nn}{NGC\,5128}

\usepackage{color}

\definecolor{darkred}{rgb}{0.625,0.,0.}

\definecolor{purple}{rgb}{0.525,0,0.525}

\definecolor{aucol}{rgb}{0.,0.5,0.3}

\defcitealias{2014MNRAS.443..969C}{CA14}

\defcitealias{2020MNRAS.497.1654C}{CS20}

\title[Probing cool giants with fluctuation eigenspectra]{Probing cool giants in unresolved galaxies using fluctuation eigenspectra:
A demonstration using 
high-resolution 
MUSE 
observations of NGC\,5128}

\author[R.J. Smith]{
Russell J. Smith$^{1}$\thanks{russell.smith@durham.ac.uk}
\\
$^{1}$Centre for Extragalactic Astronomy and Department of Physics, University of Durham, Durham DH1 3LE, United Kingdom\\
}

\date{Accepted 2021 November 18. Received 2021 November 12; in original form 2021 September 6}

\pubyear{2021}

\begin{document}

\label{firstpage}
\pagerange{\pageref{firstpage}--\pageref{lastpage}}
\maketitle

\begin{abstract}
I describe and demonstrate a new approach to using spectroscopic data to exploit Poisson sampling fluctuations in unresolved stellar populations.
The method is introduced using spectra predicted for independent samples of stars from a 10\,Gyr population a using a simple stochastic spectral synthesis model. A principal components analysis shows that $>$99 per cent of the spectral variation in the red-optical can be attributed to just three ``fluctuation eigenspectra'', which can be related to the number of giant stars present in each sample, and their distribution along the isochrone. The first eigenspectrum effectively encodes the spectrum of the coolest giant branch stars, and is equivalent to the ratio between high- and low-flux pixels discussed in previous literature.  The second and third eigenspectra carry higher-order information from which the giant-star spectral sequence can in principle be reconstructed. I demonstrate the method in practice using observations of part of NGC\,5128, obtained with the MUSE narrow-field adaptive optics mode. The expected first eigenspectrum is easily recovered from the data, and closely matches the model results except for small differences around the Ca\,{\sc ii} triplet.  The second eigenspectrum is below the noise level of the present observations. A future application of the method would be to the cores of giant ellipticals to probe the spectra of cool giant stars at high metallicity and with element abundance patterns not accessible in the Milky Way. 
\end{abstract}


\section{Introduction}\label{sec:intro}

Integrated-light spectroscopy is the pre-eminent method for studying the stellar content of passive galaxies that are too distant to be resolved into individual stars. The total observed spectrum mixes signal from (at least) cool dwarfs, warm turn-off stars, and evolved giants, in varying proportions. Yet by comparison to the equivalent  spectrum predicted from population synthesis models, a rich variety of properties can be constrained, including age or star-formation history, the initial mass function (IMF), and metallicity or individual element abundances \citep[see, e.g][]{2013ARA&A..51..393C}.
Such studies perform the twin roles of (a) probing the formation histories of galaxies beyond the reach of resolved photometry, and (b) testing the model inputs and stellar physics under conditions which differ from those pertaining in the Milky Way. The cores of elliptical galaxies are especially interesting from both perspectives: they are thought to have formed at the earliest epochs in violent starbursts, and they exhibit peculiarities, such as super-solar metallicity, enhancements in $\alpha$ elements and in sodium, and 
strong dwarf-star spectral features, attributed to a bottom heavy IMF 
\citep[e.g.][]{2014ApJ...780...33C,2017ApJ...841...68V,2019MNRAS.489.4090L}.

Although truly resolved studies of individual stars in the cores of ellipticals are impossible, additional information about the brightest stars can in principle be retrieved from high-resolution observations, via the measurable statistical fluctuations in the 
small number of giants contributing to each resolution element.
In the imaging domain, this effect has long been employed in the form of surface brightness fluctuations (SBFs) as introduced by \cite{1988AJ.....96..807T}.  The fluctuations in broad-band flux, on the scale of the image point spread function (PSF), reflect the magnitude of the brightest stars present, which dominate the Poisson variance. The SBFs can thus be used as a distance indicator, by calibrating the fluctuation magnitude as a standard candle \citep[e.g.][]{2001ApJ...546..681T} and as a probe of the stellar populations, where they carry information which is distinct from the integrated colour or spectrum \citep*[e.g.][]{2002ApJ...564..216L}.

Several works have considered the exploitation of Poisson fluctuations in the spectroscopic domain.  One approach is simply to apply the standard SBF concept to narrow wavelength channels. The ``SBF spectrum'' can be straightforwardly calculated 
from stellar population synthesis models 
\citep[e.g.][]{2018MNRAS.480..629M,2020MNRAS.493.5131V}, 
as the ratio of the variance to the mean flux at each wavelength.
In practice, the SBF spectrum can be measured in datacubes from integral field unit (IFU) spectrographs, by applying the classic imaging-processing technique on a channel-by-channel basis.
\cite{2018MNRAS.480..629M} have pioneered this approach, using 
MUSE\footnote{The Multi-Unit Spectroscopic Explorer \citep{2010SPIE.7735E..08B} on the ESO 8.2m Very Large Telescope.}
observations of the post-starburst S0 galaxy NGC\,5102. They confirmed that the observed SBF spectrum is dominated by the spectral features of M-type giant stars, and showed that using a combination of the mean and SBF spectrum could tighten the inferred constraints on the age and metallicity of the stellar population, compared to the mean spectrum alone.

An alternative approach to the Poisson fluctuation regime was discussed by
\cite{2014ApJ...797...56V}. Noting that the SBFs in imaging data are driven by variation in the number of luminous stars, they used synthesis models to compute the expected ratio of spectra from high- and low-brightness pixels. In contrast to SBF spectra, which treat each wavelength channel independently, this ``fluctuation-ratio spectrum'' explicitly captures the correlated spectral structure in the pixel-to-pixel variation. 
In \cite{2014ApJ...797...56V}, the method was demonstrated not on true spectroscopic data but instead by using the spectra to predict how narrow-band pixel colours should correlate with brightness fluctuation in images from the {\it Hubble Space Telescope (HST)}.  

In this paper, I introduce an alternative approach to spectroscopy in the fluctuation regime, based on an eigenspectrum decomposition of the variation among independent samples from the population. Section~\ref{sec:stochspec} first describes the construction of synthetic data by Poisson sampling from an underlying simple stellar population model. A principal components analysis is then applied to retrieve the leading eigenspectra, and the relationship of these to the underlying stellar content is explored.
Section~\ref{sec:obsdemo} presents a first observational application of this method, using adaptive optics observations of a stellar field in the peculiar early-type galaxy NGC\,5128. The results are evaluated through comparison to the model eigenspectra, and by inspection of the corresponding eigenvalue maps. 
 Section~\ref{sec:disc} discusses the merits of and challenges for the method, and prospects for its further application to more distant, but less disturbed, elliptical galaxies. The main conclusions are summarised in 
Section~\ref{sec:concs}.

\begin{figure*}
\includegraphics[width=177mm,angle=0]{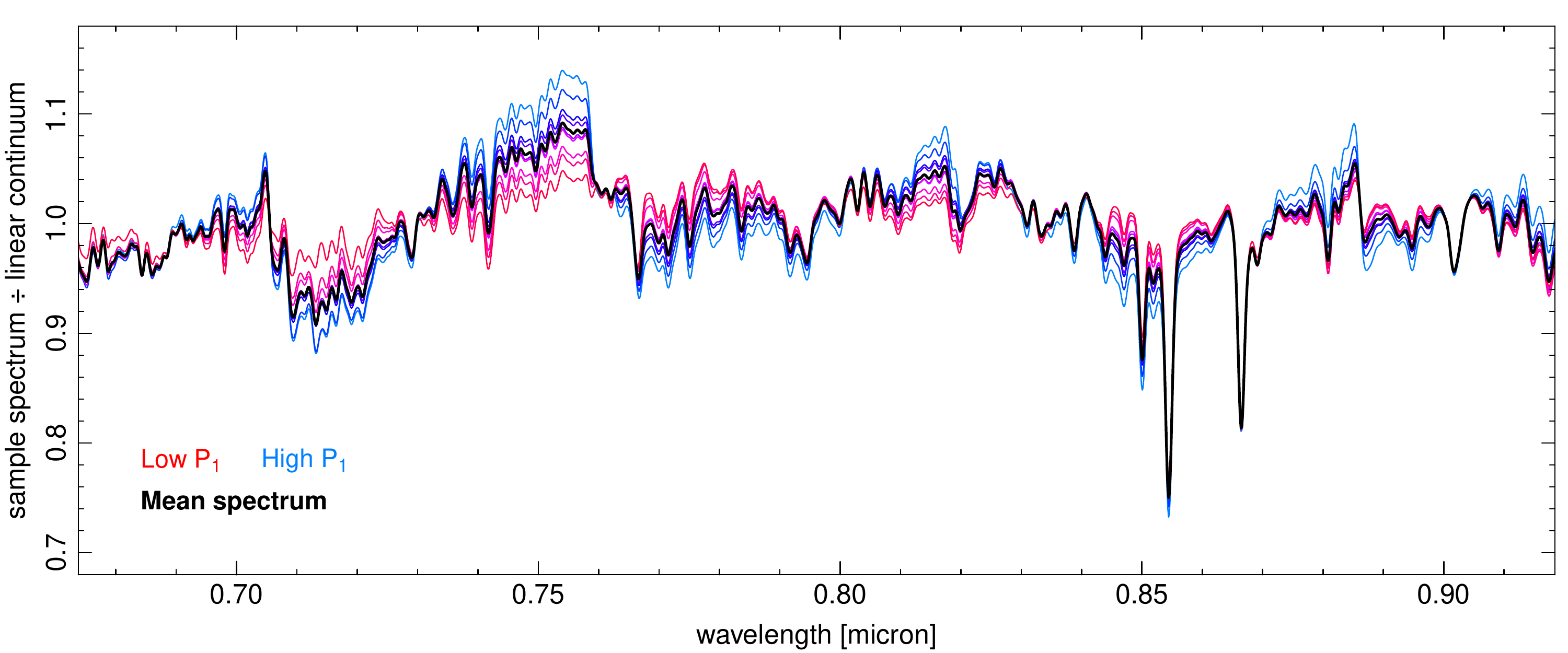}
\vskip -1mm
\caption{Ten samples from the 10\,Gyr stochastic spectral synthesis model, showing the variation caused by Poisson fluctuations in the giant stars contributing to independent pixels. The plotting colour is ordered according to the first eigenvalue, $P_1$, derived in the principal components analysis. The black line shows the mean spectrum over all 3500 samples computed, not just those shown. A linear continuum has been divided out of each spectrum.}\label{fig:tensamps}
\end{figure*}

\section{Fluctuation eigenspectra}\label{sec:stochspec}

In this section, I describe the process of building spectra from stochastically-sampled population 
synthesis models, and decomposing them into eigenspectra which describe the major contributions
to the variation among independent samples.  Some properties of the synthetic
data 
are chosen to match the characteristics of 
the MUSE data analysed in Section~\ref{sec:obsdemo}, 
but most of the principles are quite general.

The calculations are based on the usual ingredients and processes used for spectral synthesis; the only unusual feature is the inclusion of stochastic sampling from the 
mass function. 
The process begins with an isochrone describing the locus of the stellar parameters
(temperature, surface gravity, luminosity, etc), as a function of initial mass. 
All of the models in this paper are computed using the 10\,Gyr isochrone with solar metallicity from the PARSEC\footnote{PAdova and TRieste Stellar Evolution Code.} stellar tracks\footnote{These tracks include evolution on the asymptotic giant branch up to the first thermal pulse; including the subsequent evolution using the tracks of \cite{2013MNRAS.434..488M} did not affect any of the results for the old populations used here.} \citep{2012MNRAS.427..127B}.
Appropriate stellar spectra are then associated to the isochrone points, using the {\sc spi} approximator\footnote{Spectrum Polynomial Interpolator  \citep{2017ApJS..230...23V}.}
to 
interpolate between empirical spectra from the MILES\footnote{Medium-resolution Isaac Newton Telescope Library of Empirical Spectra.} \citep{2006MNRAS.371..703S} and Extended IRTF\footnote{Infrared Telescope Facility.} \citep{2017ApJS..230...23V} libraries. For ease of comparison to the data for \nn, all spectra are smoothed to 
its characteristic velocity dispersion of 150\,km\,s$^{-1}$ \citep{1986MNRAS.218..297W}.
Given the IMF (here a single power law with the Salpeter slope), the mean number of stars present on each segment of the isochrone can be defined up to a normalising constant, which depends 
on the surface brightness, the distance, 
and the angular area of the spatial sample. 
The isochone is then populated by 
drawing Poisson-distributed star counts according to these means, and the stellar contributions  are summed with the corresponding  number weights and luminosity weights, to 
generate the integrated spectrum $S_i(\lambda)$ for the $i$'th individual sample. For the illustrative purposes of this paper, the samples are assumed to be fully independent, but the effects of smoothing by a PSF could be included in more realistic calculations.

Figure~\ref{fig:tensamps} shows the red-optical spectra computed for ten samples from the population, compared to the mean synthesised spectrum.  The calculations are matched to the observational case described in Section~\ref{sec:obsdemo}.
In particular, the adopted surface brightness ($\mu_I$\,=\,
18.3\,mag\,arcsec$^{-2}$), and distance modulus (27.9\,mag) are appropriate to the observed field in \nn, the spectra are rebinned at 1.25\,\AA\ intervals (as for MUSE) and limited to 6750--9200\,\AA\ (where the observed PSF is sharpest), and the spatial samples are 0.125$\times$0.125\,arcsec$^2$ in area (5$\times$5 MUSE pixels in the narrow-field mode). With this configuration, the mean luminosity per sample is 
$\sim$4000\,$L_\odot$, of which $\sim$60 per cent is generated by $\sim$130 giant stars.
In Figure~\ref{fig:tensamps}, the structured effect of the Poisson  fluctuations in these giants is clearly visible: samples that are fainter at $\sim$0.715\,$\mu$m are typically  
brighter at $\sim$0.755\,$\mu$m and $\sim$0.815\,$\mu$m, but fainter at $\sim$0.770\,$\mu$m.  Note that a linear continuum shape has been divided out of all the spectra before plotting, to emphasise the narrower spectral features, and
to mimic the treatment of the models and data below.

The variations in Figure~\ref{fig:tensamps}  arise from the most luminous giants, which are present in small number and hence subject to the largest statistical fluctuations. 
The spectra of these cool stars are dominated by molecular band heads from the Ti\,O  $\gamma$, $\delta$ and $\varepsilon$ systems.
As described in Section~\ref{sec:intro}, the Poisson sampling variation has sometimes been described
by computing the SBF spectrum, defined as $\rm{var}(S(\lambda))/\langle{}S(\lambda){}\rangle$ \citep{2018MNRAS.480..629M}. 
This method recovers a strong signal in the Ti\,O bands because M-giants dominate the fluctuations, but since it treats each spectral channel independently, the SBF spectrum does not explicitly encode the {\it correlated} spectral variations seen in Figure~\ref{fig:tensamps}.
The  \cite{2014ApJ...797...56V} approach, based on a ratio of  spectra between samples with higher- versus lower-than-average flux, i.e. more or fewer giants, does capture the correlated variations, but requires a level of relative photometric calibration that may be challenging to achieve with IFU data.

The new direction taken in this paper is to decompose the spectral variations between samples into a small set of eigenspectra, using principal components analysis (PCA) to identify the optimal parameterisation.
In preparation for the PCA, each synthetic spectrum is divided by a low-order polynomial continuum, to remove sample-to-sample colour variations. This ensures that the extracted information is independent of luminosity or broadband colour, which may be affected by calibration uncertainties. The spectra are also divided by the mean synthesised spectrum, to isolate the variations:
$\Delta{}S_i(\lambda)\,=\,S_i(\lambda)/\langle{}S\rangle(\lambda) - 1$.

The PCA algorithm then decomposes the correlation structure of the $\Delta{}S_i(\lambda)$ into eigenspectra $E_j(\lambda)$, ordered by their contribution to the total variance, such that
\[
\Delta{}S_i(\lambda)\,\approx\, P_1 E_1(\lambda) + P_2 E_2(\lambda) + ... + P_m E_m(\lambda) 
\]
(with different values of $P_1, P_2$, etc for each $i$)
provides an acceptable approximation, 
for some small number $m$ of retained components.
In the absence of observational error, the eigenspectra are uniquely determined by the inputs chosen for the synthesis model (e.g. the isochrone, stellar library),
and are independent of the number of stars in each sample. The eigenspectra can be considered to have the same ``status'' as an output of the model as does the mean spectrum, or the SBF spectrum.  From the discussion above, we can expect that the first fluctuation eigenspectrum, $E_1$, should be driven by the varying number of cool luminous giants present in the samples, similar to the \cite{2014ApJ...797...56V} fluctuation spectrum. 
Subsequent $E_j$ can be expected to reflect the difference in distribution of the giants in each sample, and hence 
probe
the systematic trend in stellar spectra along the giant sequence.

\begin{figure*}
\includegraphics[width=177mm,angle=0]{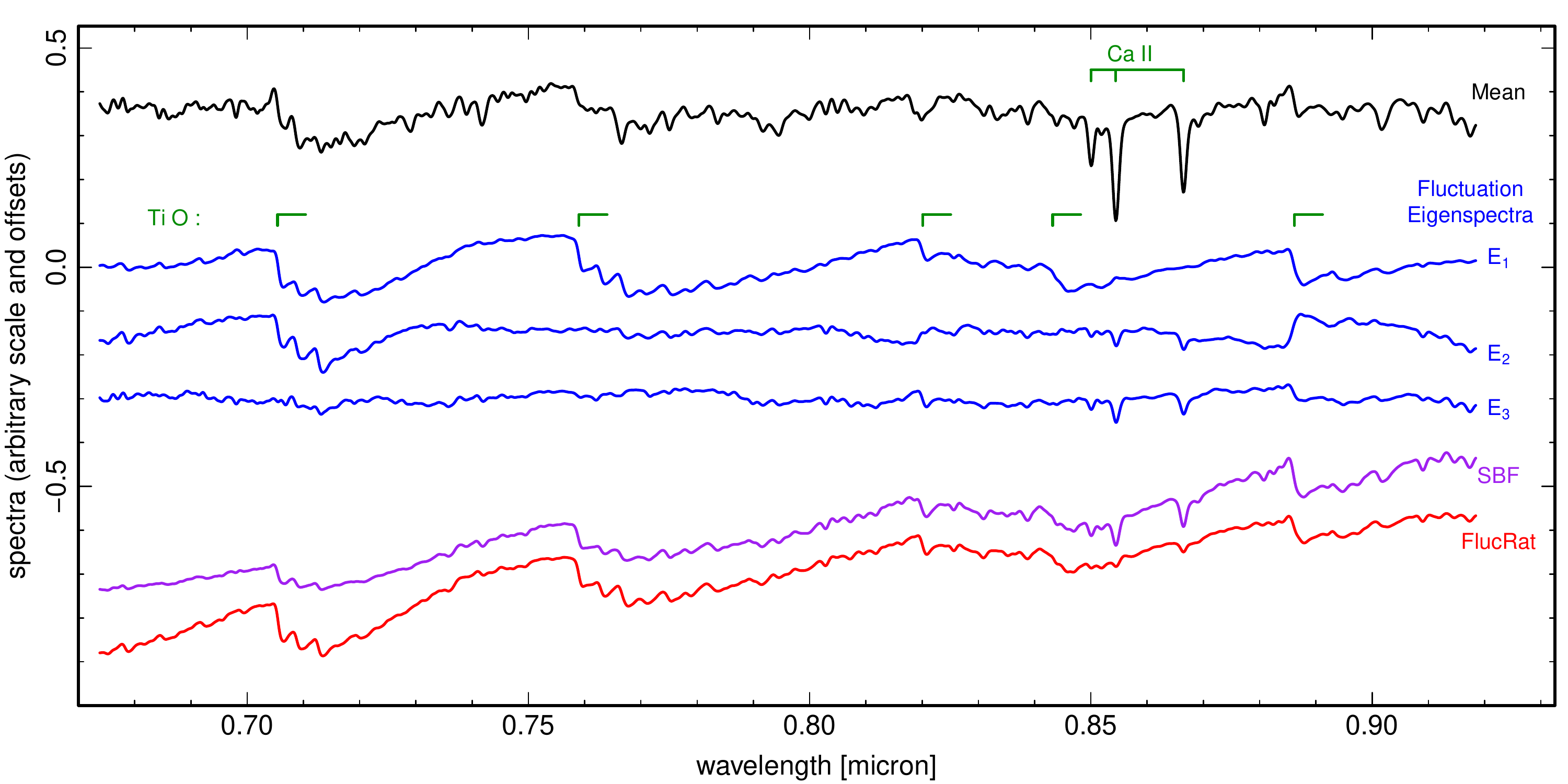}
\vskip 0mm
\caption{Fluctuation eigenspectra computed from principal component analysis of 3500 samples from the 10\,Gyr stochastic synthesis model (blue), compared to the mean spectrum (black). 
The first three components account for $>$99 per cent of the spectral variation caused by Poisson fluctuations.
For comparison, two alternative indicators of the variations are shown, computed from the same model samples: the wavelength-dependent surface brightness fluctuations as discussed, e.g. by \protect\cite{2018MNRAS.480..629M}  and \protect\cite{2020MNRAS.493.5131V} (purple), 
and the fluctuation-ratio spectrum introduced by \protect\cite{2014ApJ...797...56V} (red).
The wavelengths of some prominent Ti\,O bandheads and the Ca\,{\sc ii} triplet lines are indicated.} \label{fig:modeleigens}
\end{figure*}

\begin{figure*}
\includegraphics[width=180mm,angle=0]{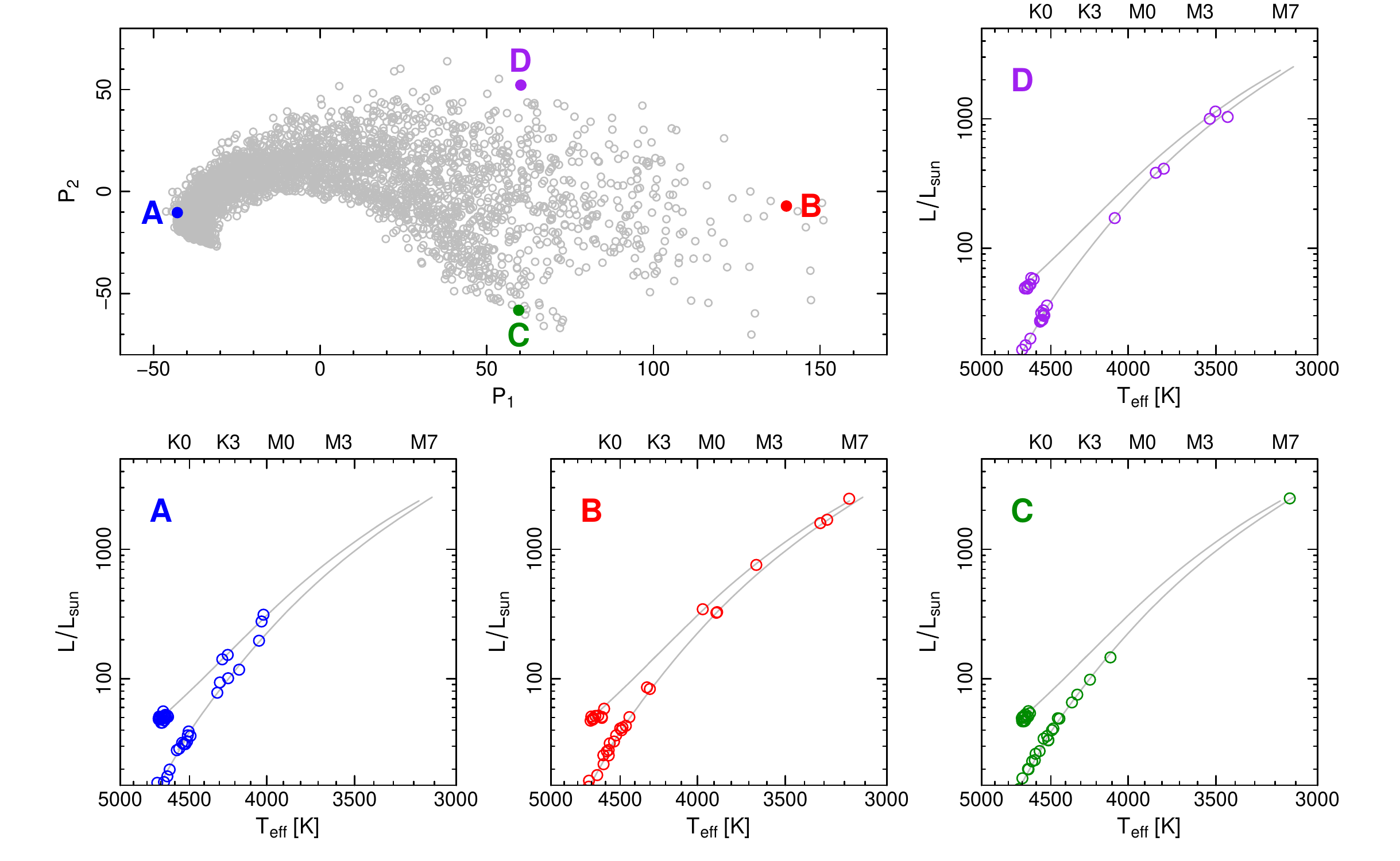}
\vskip -1mm
\caption{A visual guide to interpreting the spectral fluctuations. The upper left panel shows the first two eigenvalues, $P_1$ and $P_2$, for all of the stochastic samples computed from the 10\,Gyr stellar population. The remaining panels show the distribution of stars on the model isochrone for four samples chosen from the extremes of the eigenvalue distribution. (Small random dithers have been applied to help separate multiple stars at the same isochrone point.) Increasing $P_1$ corresponds to a larger contribution from cool giants (compare samples ``A'' and ``B''); increasing $P_2$ corresponds to a less pronounced gap in the temperature distribution (compare samples ``C'' and ``D'').
The temperature versus spectral type relation for giants is taken from \protect\cite{1999AJ....117..521V}.
}
\label{fig:pccmds}
\end{figure*}

\begin{figure*}
\includegraphics[width=170mm,angle=0]{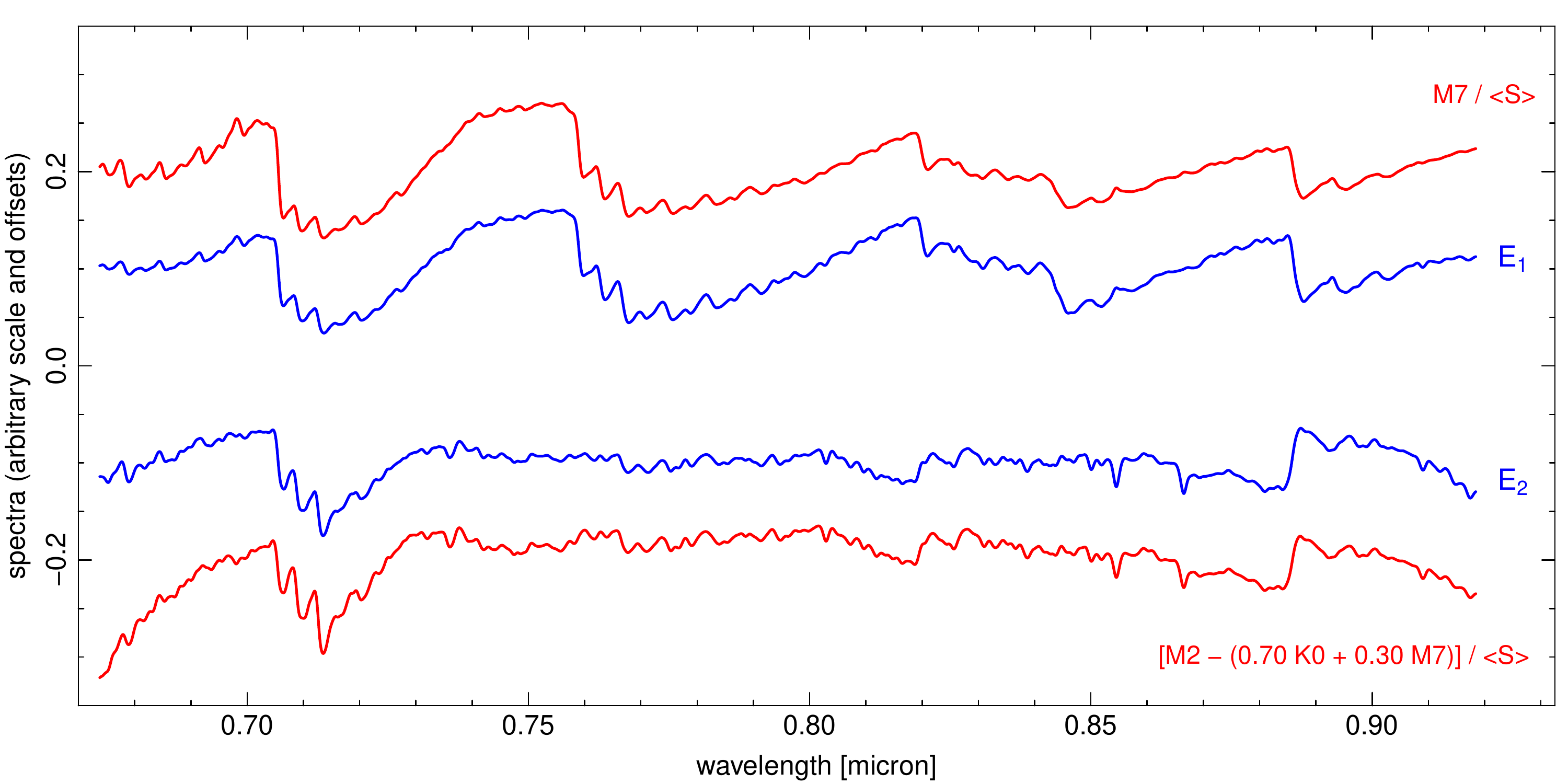}
\vskip 0mm
\caption{Relationship of the first two fluctuation eigenspectra to the spectra of individual giant stars in the model. The first eigenspectrum, $E_1$, is essentially the ratio between an M7 giant and the mean spectrum of the population.  The second, $E_2$, describes the deviation of an M2 giant from the closest approximation that can be derived from a linear combination of K0 and M7 giants, also expressed as a ratio with the mean spectrum.}
\label{fig:pcinterpret}
\end{figure*}

Figure~\ref{fig:modeleigens} shows the mean spectrum $\langle{}S\rangle$, and the eigenspectra $E_1-E_3$, computed for the 10\,Gyr-age, solar-metallicity, Salpeter-IMF population model, derived from 3500 samples similar to those in Figure~\ref{fig:tensamps}. The sign of the eigenspectra is arbitrary in the PCA process, so 
they have been plotted here such that the strong band at 0.71\,$\mu$m is shown 
in ``absorption'' in each spectrum. 

The first principal component accounts for 77.6 per cent of the total variance among samples. 
The corresponding eigenspectrum, $E_1$, is dominated by the Ti\,O bands as foreseen from Figure~\ref{fig:tensamps}, while no trace of the Ca\,{\sc ii} triplet lines is present. $E_1$ closely resembles the  \cite{2014ApJ...797...56V} fluctuation-ratio spectrum, as derived from the same model predictions.
The second principal component accounts for a further 16.2 per cent of the total variance. Its eigenspectrum, $E_2$,  again shows the strong 
Ti\,O band at 0.71\,$\mu$m, but the 
band at 0.76\,$\mu$m is barely present, and those 
at 0.82\,$\mu$m and 0.88\,$\mu$m are reversed in sign compared to $E_1$.
The  Ca\,{\sc ii} triplet lines are also more noticeable. 
Thus the second component encodes a ``correction term'', which strengthens Ca\,{\sc ii}, and
the Ti\,O bands in the blue,  but weakens those in the red.  
The third principal component accounts for a further 5.6 per cent of the total variance; its eigenspectrum, $E_3$ shows weaker molecular features, and the Ca\,{\sc ii} lines are more prominent.  All subsequent components collectively contribute only $\sim$0.6 per cent of the variance.
Thus although the stars can be distributed along the model isochrone in infinite combinations, almost all of the emergent spectral variation can be compressed into just three linear terms.

Figure~\ref{fig:pccmds} illustrates how the eigenvalues $P_1$ and $P_2$ relate to the distribution of stars on the giant branches in individual samples from the population. 
As expected from the above discussion, a larger value for $P_1$ (e.g. sample ``B'' in the figure) indeed corresponds to a larger contribution from the coolest giants, while the lowest values are found for samples having effectively no giants cooler than $\sim$\,4000\,K 
(e.g. sample ``A''). Comparing samples with similar $P_1$ values, it can be seen that higher $P_2$ corresponds to more continuously populated giant branches (e.g. sample ``D''), while lower $P_2$ is associated with a wider gap below the brightest star present (e.g. sample ``C'').

Guided by Figure~\ref{fig:pccmds}, some physical interpretation of the first two principal components can be inferred. The first component evidently encodes the total cool-giant content. Since the PCA was applied to the ratio between sample spectra and the mean spectrum, $E_1$ should be simply the ratio between a characteristic giant star and the mean. Comparison to each stellar spectrum along the isochrone reveals a close match for stars at the very tip of the giant branch, with $T_{\rm eff}$\,$=$\,3136\,K (spectral type $\sim$M7); the match to any star warmer than 3500\,K ($\sim$M4) is notably poor. 
Thus $E_1$  effectively isolates the spectrum of the coolest giant stars present in the population, which even collectively contribute only a small fraction of the total integrated light (e.g. $\sim$7 per cent for giants with $T_{\rm eff}$\,$<$\,3500\,K).

The second principal component, $E_2$, seems to encode the presence or absence of a ``gap'' on the giant branch, and so should be related to the spectrum of warmer giants than those traced by $E_1$. If the giant sequence were ``linear'', in the sense that any star could be represented as a linear combination of the M7 star with a K0 giant, for example, then the gap would not need to be described with a separate component.  It follows that $E_2$ should represent the {\it departure} of the intermediate giants from a linear spectral sequence, expressed again as a ratio with the mean spectrum. 
Minimising a least-squares statistic over the possible combinations, a very close match is found for 
$E_2\,\approx\,[{\rm M2} - ( 0.70\, {\rm K0} + 0.30\, {\rm M7})]/\langle{}S\rangle$, 
where the spectral types are shorthand for unity-normalised stellar spectra with temperatures
3671\,K, 4547\,K and 3136\,K.
Figure~\ref{fig:pcinterpret} shows how closely this approximation, and the equivalent
$E_1\,\approx\,{\rm M7} / \langle{}S\rangle$, reproduces the first two fluctuation eigenspectra.  
A comparably simple interpretation for the third principal component has not been identified. The eigenvalue $P_3$ is somewhat correlated with the number of red clump stars present, and the features in $E_3$ certainly suggest a sensitivity to higher temperatures, where the molecular bands are weaker. $E_3$ will not be discussed further here.

To summarise this section, the Poisson variation among independent samples from an old stellar population can be efficiently described using a linear combination of two or three  eigenspectra. The first two components, at least, have an intuitive meaning in terms of the spectra of the most luminous stars present in the population. The PCA approach describes the spectral fluctuations with more generality than related methods such as the SBF spectrum, and is well suited to IFU data with imperfect flatfielding and relative flux calibration.

\begin{figure*}
\includegraphics[width=178mm,angle=0]{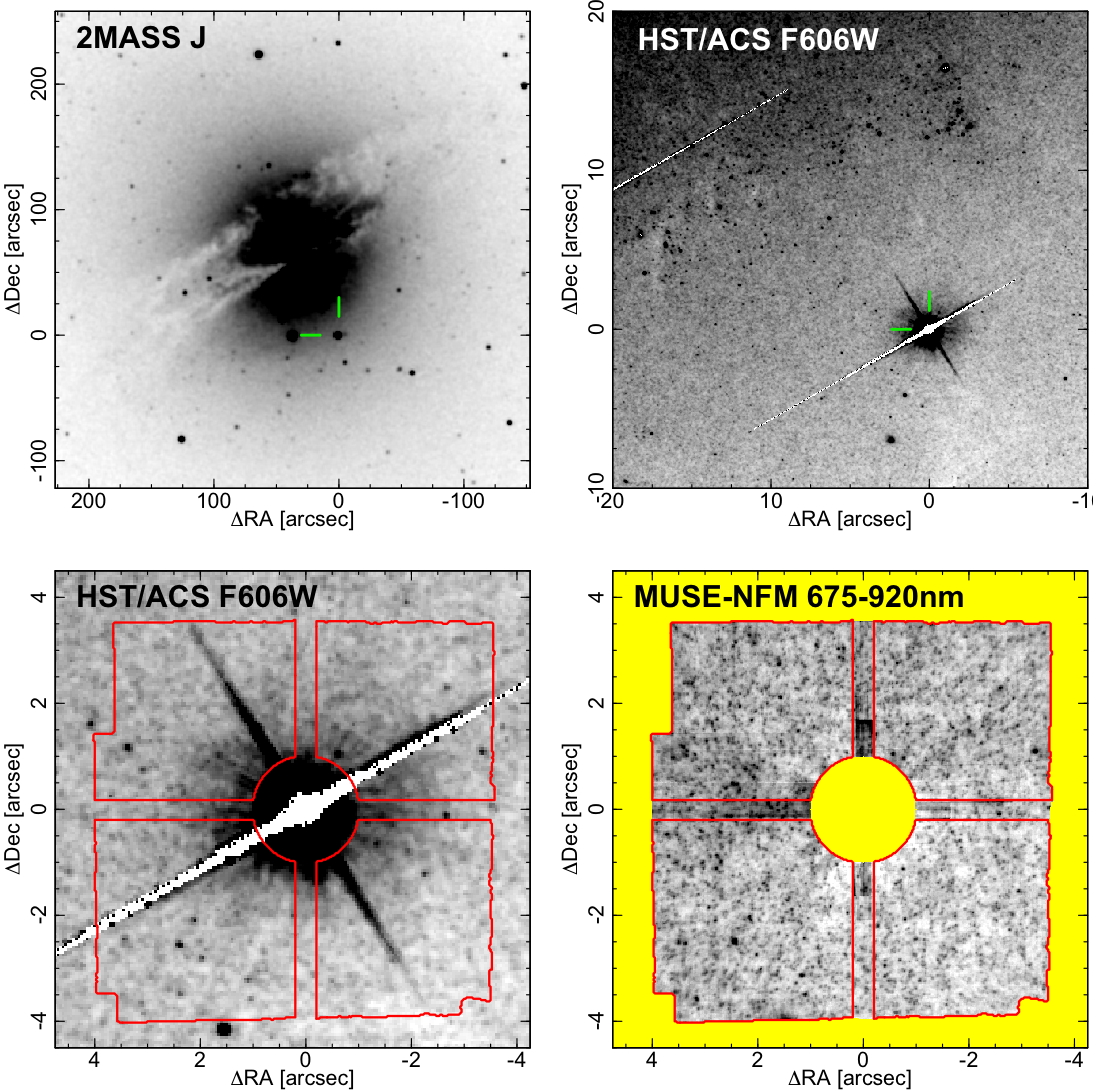}
\vskip -0mm
\caption{The NGC\,5128 MUSE AO field in  context.
The upper-left panel shows a J-band image from the 2MASS Large 
Galaxy Atlas \protect\citep{2003AJ....125..525J}, highlighting the foreground reference star used for the low-order wavefront correction. 
The upper-right panel shows an extract from {\it HST} imaging in F606W from
\protect\cite{2006AJ....132.2187H}. The targeted field  
lies away from the region of recent star formation seen to the north, and is free from prominent dust features. The lower-left panel shows a zoom of the {\it HST} image, while the lower-right shows the
collapsed red-optical image from MUSE on the same scale. The red outline shows the default mask region; pixels in the ``cross'' are affected by scattered light from the AO star, but are retained here for comparison to the {\it HST} image.  
}
\label{fig:matchedimages}
\end{figure*}

\section{Observational demonstration}\label{sec:obsdemo}

In this section I present a practical demonstration of  extracting the fluctuation eigenspectra from IFU data, using new high-resolution adaptive optics (AO) observations of a field in NGC\,5128.

The data were acquired using MUSE at the ESO Very Large Telescope, operated in its narrow-field mode, which implements Laser Tomographic Adaptive Optics using four laser  beacons \citep{2012SPIE.8447E..37S}.  The instrument 
provides a $\sim$7\,arcsec field-of-view, sampled with 0.025\,arcsec spatial pixels.
NGC\,5128 was chosen as the target for this observation mainly for its proximity 
\citep*[3.8$\pm$0.1\,Mpc; ][]{2010PASA...27..457H}, 
since the Poisson fluctuations are larger for closer galaxies, at 
a given surface brightness.
A bright on-axis guide star was required for the tip/tilt 
correction in the narrow-field mode when these observations were proposed, 
and so a field was selected at $\sim$80\,arcsec ($\sim$1.5\,kpc) 
from the galaxy nucleus, centred on a $H$\,=\,11.0 foreground star (see top left panel of Fig.~\ref{fig:matchedimages}). {\it HST} imaging shows that this field is largely free 
from dust and from bright young stars (top right panel).
The surface brightness at this radius is $\mu_J$\,=\,17.4\,mag\,arcsec$^{-2}$
\citep{2003AJ....125..525J}, which converts to 
$\mu_I$\,$\approx$\,18.3\,mag\,arcsec$^{-2}$ 
for an old stellar population.

The observations were made between Feb 2020 and Apr 2021, and comprise four on-target exposures with total integration time 7122\,sec. Each individual $\sim$1800\,sec exposure was bracketed by two $\sim$330\,sec offset sky frames. (One further exposure was taken but not used in the analysis as the image quality was much poorer.) Three of the frames have approximately the same pointing, while the fourth was unintentionally offset  $\sim$2\,arcsec south; each exposure was acquired at a different position angle, with relative rotations of 90 degrees.

\begin{figure}
\includegraphics[width=83mm,angle=0]{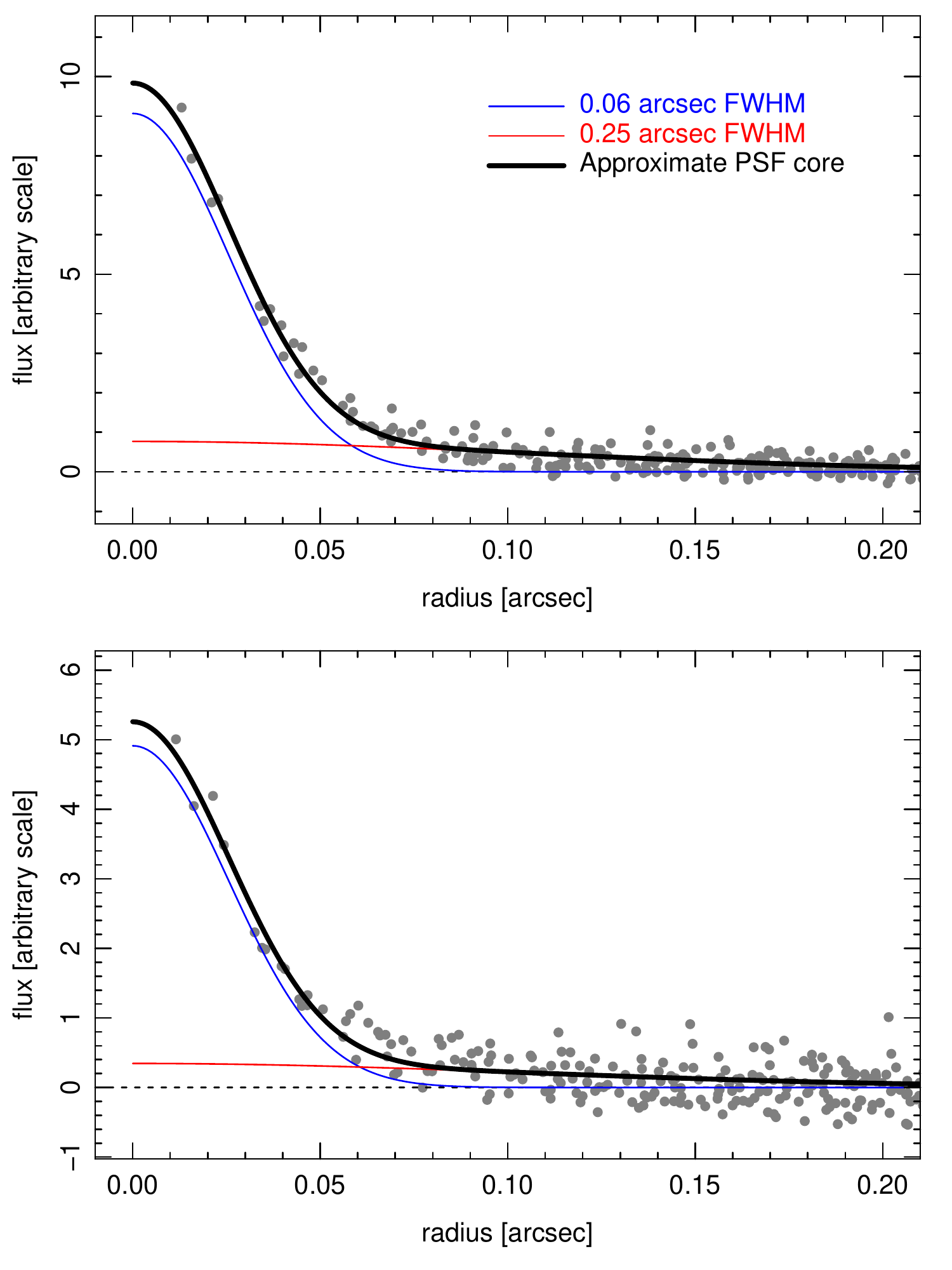}
\vskip -1mm
\caption{The core of the AO PSF as traced by two faint stars
that were not used in the frame registration process.
The upper panel shows the $I$\,$\approx$\,21.8 star seen at (+2\farcs5, --2\farcs5) 
in Figure~\ref{fig:matchedimages}, while the lower panel is the for the  $I$\,$\approx$\,22.4 star at (--3\farcs3, +1\farcs1)\,arcsec. The inner profile can be adequately described with a double-Gaussian fit as shown, though a significant outer halo is also present.}\label{fig:psfplot}
\end{figure}

The custom analysis began from the four observatory-processed pipeline-reduced and sky-subtracted data cubes. 
The frames were registered with nearest-pixel shifts determined from the image centroid of the AO star, and the cubes were combined with a median, to clean residual cosmetic artifacts. 
Only the area covered by at least three pointings was retained for further analysis.
Figure~\ref{fig:psfplot} shows the profile of the PSF core in the resulting combined frame, for two faint stars located $\sim$3.5\,arcsec from the reference star, and not used in the registration process. These profiles can be described by a Gaussian with 0.06\,arcsec full width at half maximum (FWHM), combined with a second component of 0.25\,arcsec FWHM. 
Beyond this well-corrected core, the PSF has a diffuse halo which can be traced  to  $\sim$2\,arcsec in the bright star. 
A square
5$\times$5 
(or 9$\times$9)
pixel block centred on the reference star encloses 
38.2  (or 50.4) 
per cent of the total energy in the PSF.

The halo of the reference star must be subtracted from the data
to prevent its spatially-varying light contribution from affecting the PCA fit.  It proved too difficult to model out the star based on its distinct spectrum compared to the galaxy, so the method adopted instead assumes a circular spatial profile.  
All pixels within 1\,arcsec of the star were first masked out, and 
a polynomial approximation to the wings of the 
radial profile was then fit and subtracted independently for each spectral channel.
Forming a collapsed image over the red wavelength region of the subtracted cube
confirms detection of numerous faint stars, a few of which can be matched to sources in the {\it HST} image (lower panels of Fig.~\ref{fig:matchedimages}). 
The brightest stars seen in the MUSE image  have $I$\,$\approx$\,22, while the 
typical fainter stars have $I$\,$\approx$\,24, consistent with being red giants  in \nn.
Note however, that the image combines signal over nearly 2000 wavelength channels; very few stars are detectable in the individual channels\footnote{Only the star at (+2\farcs5, --2\farcs5), with $I$\,$\approx$\,21.8, is bright enough to extract an informative spectrum. The radial velocity  confirms this star as a member of \nn, 
two magnitudes above the tip of the red giant branch.}.

The collapsed image reveals some residual scattered light in a cross pattern around the bright star, which traces the MUSE field-splitter boundaries. 
These regions are masked out from the PCA analysis.
A region of diffuse nebular emission (H$\beta$, H$\alpha$, [N\,{\sc ii}], etc) is present in 
the north-east corner of the field-of-view, i.e. towards the disk of \nn; this region was not masked for the default analysis, since the fluctuation analysis is restricted to redder wavelengths, where no strong emission lines are present.

The unmasked pixels were combined into individual spatial samples by median-binning over 5$\times$5 pixel (0.125\,$\times$\,0.125\,arcsec$^2$) blocks, chosen to roughly match the resolution of the PSF core. (Repeating the analysis with 9$\times$9 pixel blocks did not qualitatively alter the results.) For the default masking scheme described above, this yields 2490 spectra, each of which is individually extremely noisy, 
with no recognisable features.
Due to the method adopted to remove the AO star contamination, the bin spectra do not have a correct continuum level, since a mean was already subtracted at each radius for each channel. Thus for the observed data, 
the PCA was applied not to
$\Delta S_i(\lambda)$ 
as defined in Section~\ref{sec:stochspec}, but simply to 
$S_i(\lambda) - \langle S_i(\lambda) \rangle$.
The wavelength range was restricted to 6750--9200\,\AA, since the AO correction is more effective
in the red.  The particular choice of blue limit 
was made to exclude the [S\,{\sc ii}] lines in the area with nebular emission, while the red limit avoids any contamination from faint second-order light from the AO star. 
Channels in the interval 7590--7670\,\AA\ were also excised, to avoid systematic residuals from 
imperfect correction of the atmospheric A-band in the pipeline reduction.
As a final step before applying the PCA, a low-order polynomial fit was subtracted from each spectral sample, to remove any sensitivity to broadband flux response. Again, this treatment differs from that applied to the models in Sections~\ref{sec:stochspec}, where the continuum fit was divided out.

\begin{figure*}
\includegraphics[width=177mm,angle=0]{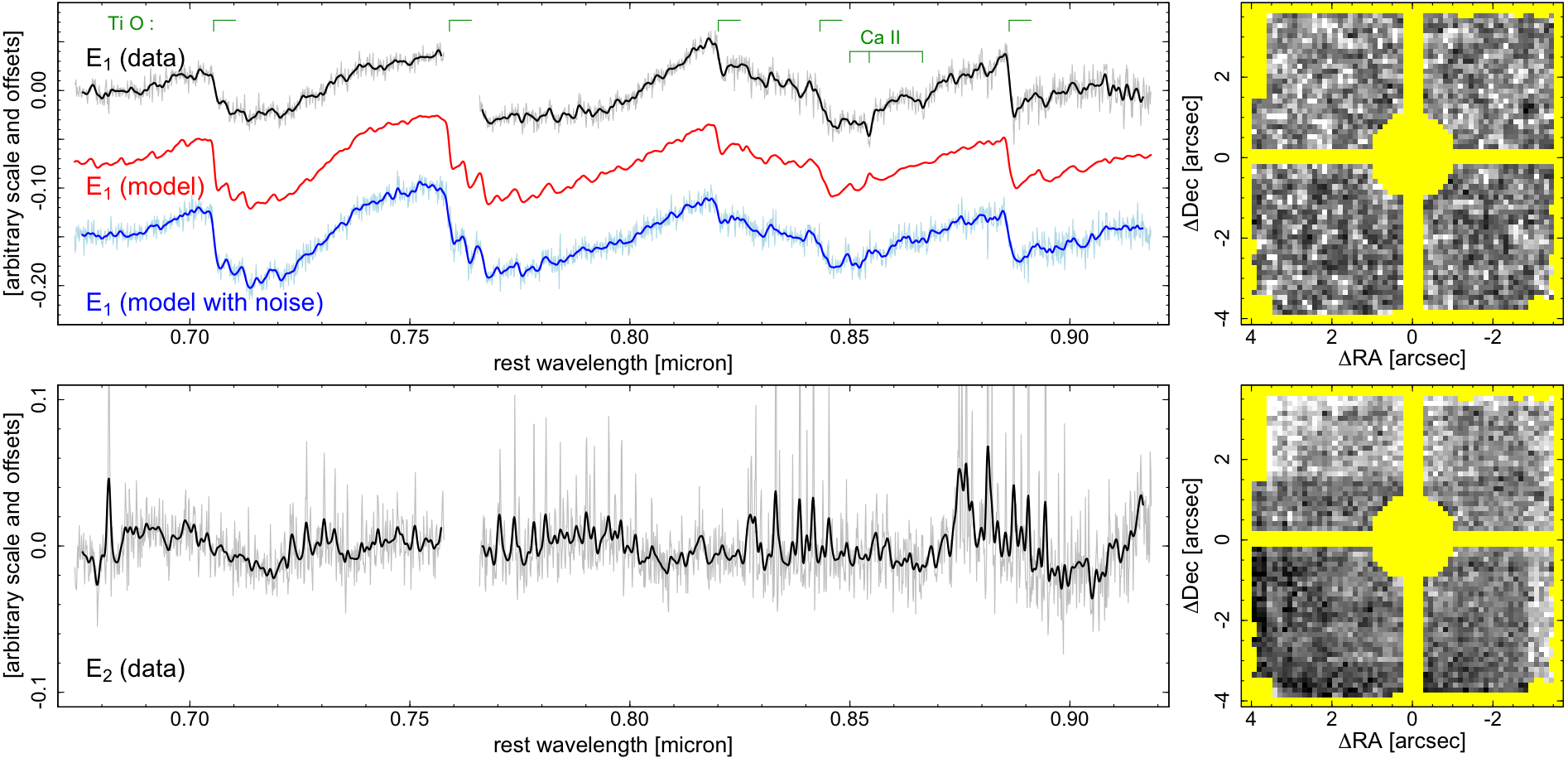}
\vskip -1mm
\caption{First and second fluctuation eigenspectra (left) and corresponding eigenvalue maps (right), derived from the \nn\ data using the default masking scheme. In the spectrum plots, the grey trace is the recovered eigenspectrum, while the black shows the same after smoothing by $\sigma$\,=\,150\,km\,s$^{-1}$ to suppress noise below the effective resolution. The  red trace for $E_1$ is the equivalent computed from the noise-free 10\,Gyr solar-metallicity stellar population model; the blue trace shows  results from the same model after adding noise to match the data quality, and smoothing as for the data. (The model PCA has here been recomputed to match the slightly different treatment necessary for the data; see text for details.)
The eigenvalue map shows that the variations associated with $E_1$ are smoothly distributed across the field-of-view, as expected if $P_1$ traces stochastic fluctuations in the giant-star content per spatial sample. The second eigenspectrum shows sharp features due to variations in the sky-subtraction residuals, and the coherent structure in its eigenvalue map confirms that this signal is {\it not} astrophysical in origin.
}\label{fig:datapca1}
\end{figure*}

\begin{figure*}
\includegraphics[width=177mm,angle=0]{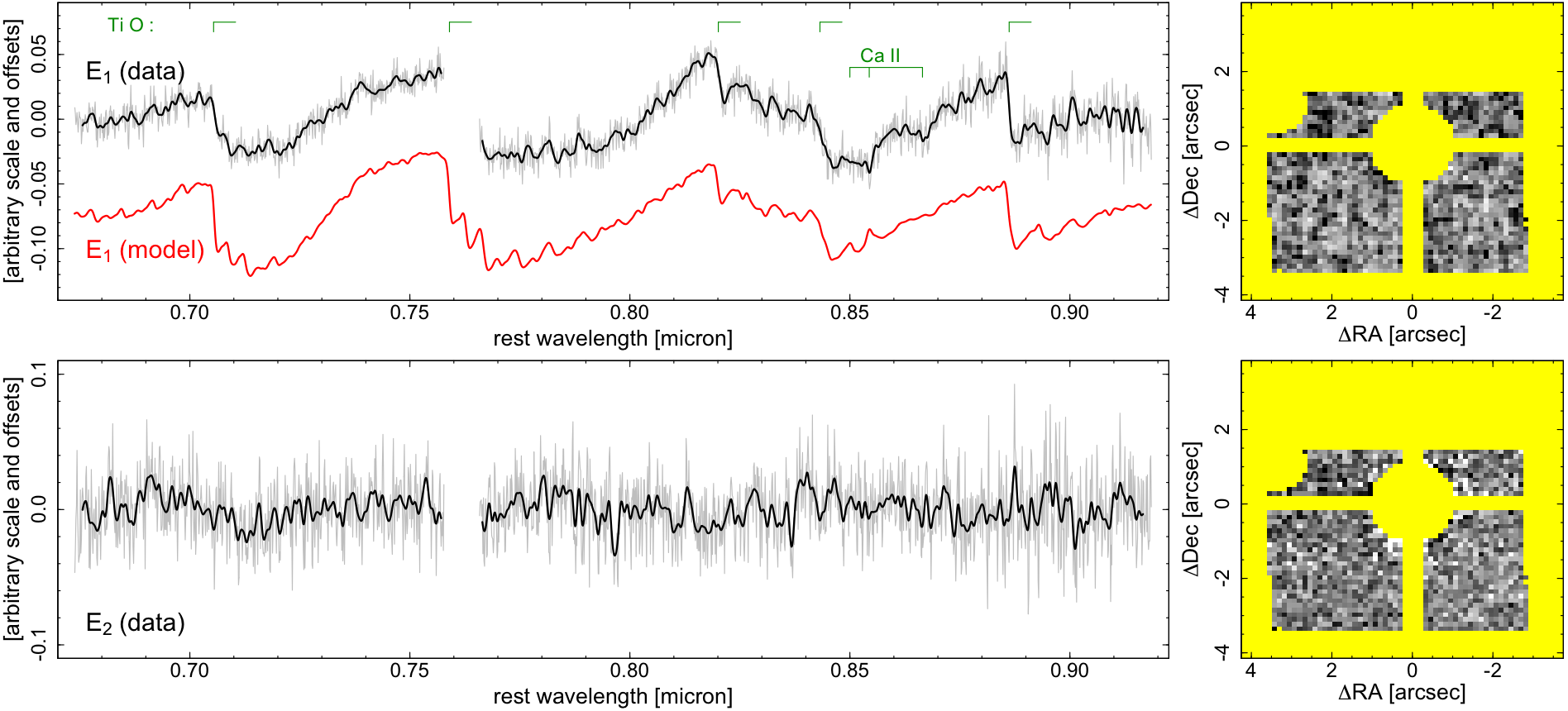}
\caption{First and second fluctuation eigenspectra retrieved using a more conservative masking scheme. The first component is almost unchanged from the default result. The spatial structure in $P_2$ is removed, but the corresponding eigenspectrum is now simply dominated by random noise.}\label{fig:datapca2}
\end{figure*}

The first two fluctuation eigenspectra recovered from the \nn\ data are shown in Figure~\ref{fig:datapca1}. 
For comparison to these results, the model predictions were regenerated to match the slightly different treatment used for the data, i.e. with subtraction of the mean and the continuum from each sample. Additionally, the PCA was run on a version of the model samples to which uncorrelated random noise was added to match the wavelength-dependent signal-to-noise ratio of the data.
Recall, however, that the model is not a fit to the  data, and has not been tuned to the specific population characteristics of \nn.

The first eigenspectrum is extracted cleanly from the observed spectra, and is remarkably free from sky subtraction artifacts.
The strong molecular band features seen in the predicted $E_1$ are immediately recognisable in the observational equivalent, and the detailed structures of the bands are very well reproduced, after applying a smoothing to $\sigma$\,=\,150\,km\,s$^{-1}$ to match the intrinsic velocity dispersion.
On the other hand, the overall relative strengths of the features differ between the observations and the model, with Ti\,O bands in the blue appearing weaker in the data compared to those in the red.
This might be astrophysical, but could also arise from the wavelength-dependent AO PSF, which leads to the spectral samples being less fully independent in the blue than in the red, which 
was not accounted for in the model calculations.
A more notable difference between the observed and model $E_1$ is apparent in the 
Ca\,{\sc ii} triplet lines, which are more prominent in the data than in the model. 
Intuitively, this suggests the presence of warmer stars among the luminous giants which dominate the fluctuation signal. A future exploration of the model parameter space should help to interpret this result.

While the fluctuation eigenspectra $E_1, E_2$, etc,  carry meaningful information about the spectra of the giant stars, the values of $P_1, P_2$ etc measured for each spatial sample
should be essentially random,
if the samples are truly drawn from the same parent population. To test whether this is the case, the right-hand panels of Figure~\ref{fig:datapca1} show the derived eigenvalues mapped across the observed field-of-view. Any coherent structure here would suggest either residual systematic errors in the data (e.g. a radial trend might reveal incomplete subtraction of the AO star light) or true astrophysical variation across the field-of-view (e.g. if the fluctuations were driven by areas with younger populations).
The map for the first eigenvalue, $P_1$, in fact does show a uniform distribution within the field, as expected 
for 
stochastic sampling variation.
As further confirmation of this behaviour, applying the PCA analysis independently for each `quadrant' of the observed field yields $E_1$ spectra consistent with the whole-field results.

The observed second eigenspectrum and its corresponding eigenvalue map indicate that this component is {\it not} recovering the variation in giant-star distribution encoded in the model $E_2$. Strong features are visible in the spectrum around the major OH sky line complexes, and a distinct boundary is present in the $P_2$ map, dividing regions observed in four exposures from those with only three. 
Evidently $E_2$ is being driven by variations in the sky-subtraction residuals between the four frames. 
Figure~\ref{fig:datapca2} shows the equivalent results after applying a more conservative masking of the data, which retains only pixels covered by all four exposures, and also excises the region where extended ionised gas was detected. 
While $E_1$ is essentially unchanged by this more restrictive mask, the second component no longer shows any spatial structure in the eigenvalue map, or obvious sky features in the eigenspectrum.
However, except for a possible weak dip at 0.715\,$\mu$m, the recovered $E_2$ does not show any clear signal from the stellar features seen in the models.

\begin{figure}
\includegraphics[width=85mm,angle=0]{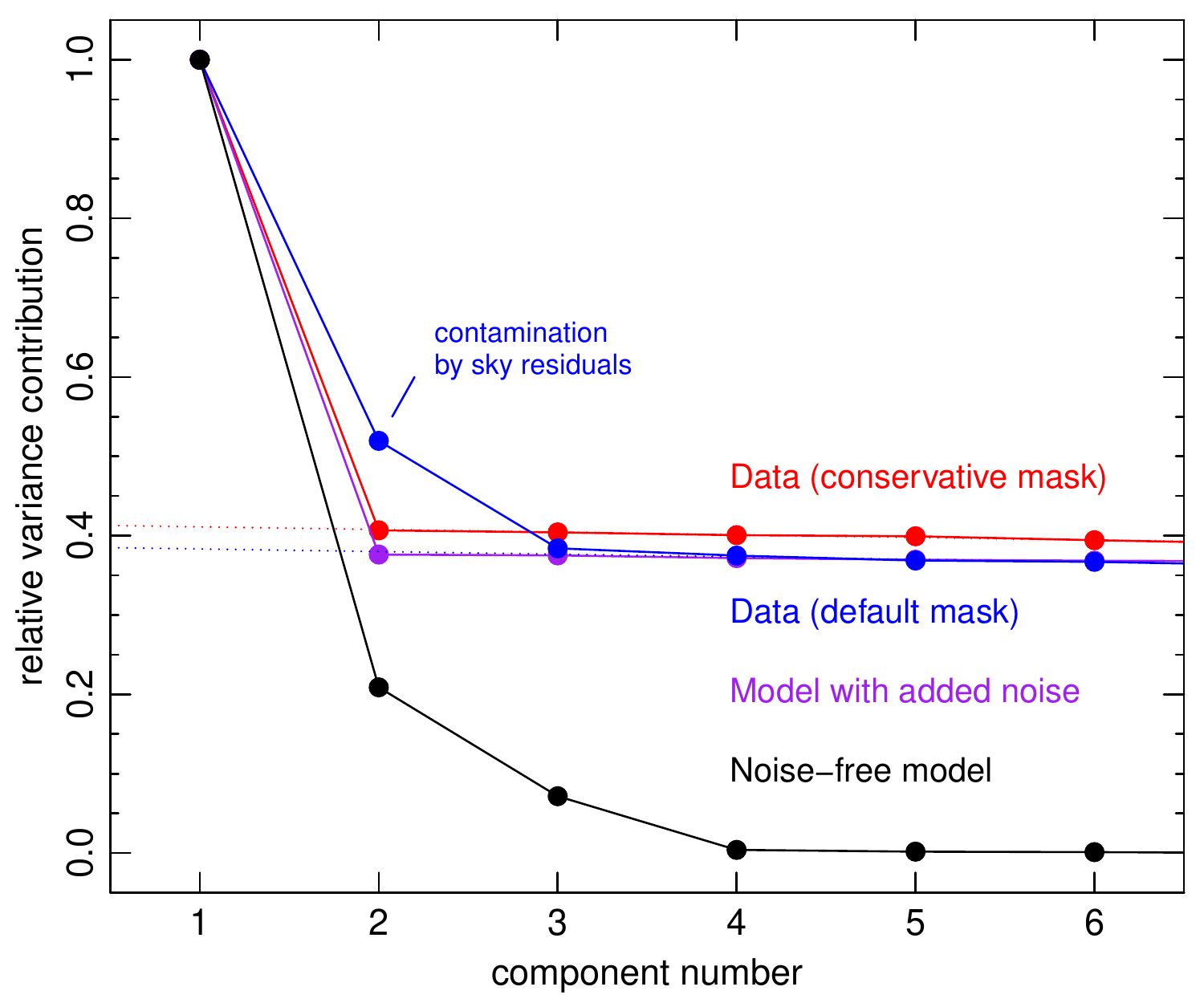}
\vskip -1mm
\caption{Total variance associated with each principal component, relative to that of the leading component. 
The noise floor of the \nn\ data (where the variance contributions flatten) is a factor of $\sim$2 higher than the expected level of the second component.  The dotted lines here show a linear fit to components 4--10 for the observations. 
}\label{fig:screeplot}
\end{figure}

Applying the PCA algorithm to the noisy observational data (typical  $S/N$\,$\approx$\,2 per sample), the leading components naturally account for a very much smaller fraction of the total variance than in the noise-free simulations. For the default masking scheme, $E_1$ contributes 0.51 per cent, while $E_2$ carries 0.26 per cent and many subsequent components then contribute at an almost constant level of $\sim$0.19 per cent. (The hundred-fold reduction in variance for the lead component is reproduced in the model PCA, when noise is added to match the data quality.) 
With the more conservative mask, $E_1$ contributes 0.63 per cent of the variance, while $E_2$ and subsequent components each carry $\sim$0.25 per cent.
Figure~\ref{fig:screeplot} compares these variance contributions between the model and the observations, normalising to 
$E_1$ in each case, to assess whether a reliable measurement of $E_2$ is feasible. Apart from the sky-contaminated component in the  default-mask results, the observational
curves flatten at a level around twice as high as the expected intrinsic variation from the second eigenspectrum, showing that substantially deeper data would be necessary to 
retrieve $E_2$ from this field.

To summarize this section, the MUSE observations of \nn\ provide enough information to 
extract the spectrum of the most luminous giants, as characterised by $E_1$, but the second-order
information carried by $E_2$ lies below the noise level of the present data.

\section{Discussion}\label{sec:disc}

The method outlined in this paper provides a general framework to extract the spectroscopic properties of the brightest stars in an unresolved stellar population. 
Applied to the red-optical spectrum of an old population, the fluctuation eigenspectra probe  cool giants on the first and early asymptotic giant branch (AGB)\footnote{Resolved spectroscopy of such stars beyond the Local Group, and in NGC\,5128 in particular, has been advanced as a key science case for instruments on 30m-class telescopes \citep{2018arXiv181004422G}.}.
However in principle the method could be applied to younger populations, where  thermally-pulsing AGB stars or red supergiants might drive the fluctuations in the red or infra-red.  If applied in the blue (notwithstanding the difficulty of AO observations at such wavelengths), hot horizontal-branch stars might
contribute significant fluctuations in the Balmer lines, if present. 

Compared to other methods proposed for the Poisson fluctuation regime \citep{2014ApJ...797...56V,2018MNRAS.480..629M}, the PCA approach has the advantage of capturing more than just a single spectral component, and hence reconstructing the stellar content in greater detail, if data quality allows. 
From a practical perspective, it is also advantageous in bypassing any dependence on broadband throughput and wavelength response. 
An inherent disadvantage of the method, however, is that the eigenspectra, and their ordering, are not guaranteed to be stable as the configuration of the data change. For example, extending the wavelength range might bring in a new spectral feature such that a different type of star begins to contribute to, or even dominate, the variance. 
Interpreting the observations thus requires new predictions  to be generated for each new observational configuration.

In this proof-of-concept paper, I have compared the observational results only to the predictions from a single fiducial stellar population model, without exploring the dependence on age and metallicity, composite populations, or the specific choice of isochrones etc; all such questions are deferred to future work. For \nn\ in particular, it will be possible to test whether the difference in $E_1$ around the Ca\,{\sc ii} triplet can be resolved by including a broad metallicity distribution, as indicated by resolved photometry \citep{1996ApJ...465...79S,1999AJ....117..855H}, or AGB stars from a  younger component \citep{2021arXiv211102945R}.
More realistic modelling should also take into account the wavelength-dependent PSF, which will introduce a systematic variation in the fluctuation strength across the spectral range.

The ultimate application of the method would be to determine the properties of cool giant stars not in a system like \nn, where they probably closely resemble those in the Milky Way, but in the centres of very massive elliptical galaxies. The stellar populations in such regions have super-solar metallicities and enhanced [$\alpha$/Fe] and [Na/Fe] ratios \citep*{2014ApJ...780...33C}, as well as other spectral signatures that seem to indicate an IMF enriched in 
low-mass stars \citep{2020ARA&A..58..577S}.  The integrated spectra of elliptical galaxies are analysed through sophisticated models built usually from empirical libraries of Milky Way stars, and augmented with theoretical stellar atmosphere calculations to account for non-solar abundance patterns \citep[e.g.][]{2012ApJ...747...69C,2017MNRAS.464.3597L}.
Such calculations are especially complex for cool atmospheres with strong molecular features, and the evolutionary tracks for AGB stars also remain challenging to compute \citep{2013MNRAS.434..488M}.
Empirically isolating the spectral contribution of giants would thus provide a novel test of some of the uncertain ingredients in stellar population  models for elliptical galaxies, which cannot be performed directly in the Milky Way.

Perhaps counter-intuitively, the spectral fluctuation method is not limited to the situation seen in the \nn\ observations, where the number of luminous giants per spectral sample is of order unity. At a given distance, working at higher surface brightness reduces the relative amplitude of the fluctuations, but suppresses the photon-noise contribution to the error by the same factor, so the signal-to-noise is independent of surface brightness in this regime.  In practice, for the very small spatial pixels used in the narrow-field mode, the read noise is generally significant or dominant (as it is for the \nn\ data), so there is substantial scope to improve precision by working in brighter regions\footnote{For a population of identical stars each contributing $f$ counts, with a mean of $n$ stars per element, the intrinsic fluctuations are simply $\sqrt{n}\,f$, and the error on each measurement is $\sqrt{nf+\sigma^2}$ where $\sigma$ is the read noise. Hence the signal-to-noise increases with surface brightness at low $n$ and then flattens 
when the photon-noise regime is reached.}.
The choice of field (and consequent low surface brightness) in Section~\ref{sec:obsdemo} was driven by the need for a bright AO reference star in  early observations with the MUSE narrow-field mode. 
However, after recent upgrades to the wavefront sensor system\footnote{\url{https://www.eso.org/sci/publications/announcements/sciann17429.html}}, the low-order correction can be now achieved with point sources as faint as $J$\,$\approx$\,19 or even using the core of the target galaxy itself.
With these advances, 
retrieving at least the first fluctuation eigenspectrum in the cores of ellipticals in the Virgo or Fornax clusters might be feasible with deep MUSE observations.

\section{Summary}\label{sec:concs}

I have described a new  approach to characterising the spatial variation
in spectra caused by Poisson fluctuations in the population of bright giants sampled by each resolution element in IFU observations of galaxies.
The key results are as follows:
\begin{itemize}
    \item For old stellar population models, $>$99 per cent of the variation in the red-optical region can be characterised by a linear combination of just three eigenspectra that can be extracted through a standard principal components analysis.
    \item The first two principal components are driven by the variation in the number of cool giant stars present, and their distribution along the isochrone, in each spatial sample.  
    \item For an old stellar population, the first fluctuation eigenspectrum, $E_1$, describes the 
    coolest (late-M) giants, while the second, $E_2$, can be used to 
    reconstruct 
    the spectrum of warmer (early-M) stars.
    \item  Narrow-field mode observations of \nn\ with MUSE are able to retrieve $E_1$ with high precision; the observed features are a good, but not perfect, match to predictions from a 10\,Gyr solar-metallicity model. 
    \item Differences in $E_1$ 
    around the Ca\,{\sc ii} triplet lines suggest the presence of warmer giants stars in \nn\ than in the fiducial model.
    \item The higher order information encoded in $E_2$ lies below the noise level of the present \nn\ data.
\end{itemize}

Future application of this technique to the centres of the nearest massive ellipticals should be feasible, with recent upgrades to the MUSE AO system.
Looking further ahead, future facilities such as the IFU mode of MAVIS\footnote{Multi-conjugate adaptive optics Assisted Visible Imager and Spectrograph.} \citep{MAVISconcept} at the VLT, and HARMONI\footnote{High Angular Resolution Monolithic Optical and Near-infrared Integral field spectrograph.} \citep{2021Msngr.182....7T} at the ESO Extremely Large Telescope, offer great promise for further
such studies in the optical and infrared respectively.
 Such observations would directly probe the properties of cool giant stars with element abundance patterns not found in the Milky Way, and hence test key ingredients of current spectral synthesis models for old stellar populations.

\section*{Acknowledgements}
I am grateful for helpful comments and suggestions from the anonymous reviewer. This work was supported by the Science and Technology Facilities Council through the Durham Astronomy Consolidated Grant 2020--2023 (ST/T000244/1). 
This research is based on observations collected at the European Organisation for Astronomical Research in the Southern Hemisphere under ESO programme 0104.B-0174(A).

\section*{Data availability}
The observational data used in this paper are publicly available in the ESO archive (\url{archive.eso.org}). Other products can be provided upon reasonable request to the author.

\bibliographystyle{mnras}
\bibliography{stochspec}

\bsp	
\label{lastpage}
\end{document}